\documentclass[]{iopart} 
\begin{document}

\def\real{{\mathbf{R}}}
\def\complex{{\mathbf{C}}}
\def\wPm{{$\widehat P$-matrix}}
\def\wPms{{$\widehat P$-matrices}}

\title[$\widehat P$-matrices in orbit spaces and invariant theory]{$\widehat P$-matrices
in orbit spaces and invariant theory}

\author{V Talamini}

\address{INFN, Sezione di Padova, via Marzolo 8, I-35131 Padova, Italy}
\ead{vittorino.talamini@pd.infn.it}
\begin{abstract}
In many physical problems or applications one has to study
functions that are invariant under the action of a symmetry group
$G$ and this is best done in the orbit space of $G$ if one knows
the equations and inequalities defining the orbit space and its
strata. It is reviewed how the $\widehat P$-matrix is defined in
terms of an integrity basis and how it can be used to determine
the equations and inequalities defining the orbit space and its
strata. It is shown that the $\widehat P$-matrix is a useful tool
of constructive invariant theory, in fact, when the integrity
basis is only partially known, calculating the $\widehat P$-matrix
elements, one is able to determine the integrity basis completely.
\\

\noindent Published in {\it Journal of Physics: Conference Series}
{\bf 30} (2006) 30--40, \copyright\ copyright (2006) IOP
Publishing Ltd.
\end{abstract}


\section{Introduction}\label{Intro}
In many physical applications one has to do with functions that
are invariant under transformations of a compact symmetry group.
These functions may be considered as functions defined on the
orbit space of the group and some arguments, like the study of
symmetry breaking or the study of phase transitions, are better
understood if studied in the orbit space. To study invariant
functions in the orbit space one has to know first the equations
and inequalities that define the orbit space and its strata. A
standard way to find these equations and inequalities is by using
an integrity basis and the \wPm. This method is reviewed in
sections \ref{OS} and \ref{PmInOS}. The orbit spaces have not been
widely employed in the study of invariant functions yet because
the integrity basis is often unknown (the determination of an
integrity basis of a general group is still an open problem of
constructive invariant theory). Section \ref{Invariant} reviews
some selected arguments of invariant theory and section
\ref{PmInIT} (the main section of this article) shows how the
\wPm\ can be used together with the Molien function to determine
the integrity basis of a compact group.

\section{Orbit Spaces}\label{OS}
In this article $G$ is a compact group acting effectively in a
finite dimensional space. Then, in all generality $G$ is a group
of real orthogonal matrices acting on the real vector space
$\real^n$.

The {\it orbit} through a point $ x\in\real^n$ is the set:
$\Omega(x) = \{g \cdot x, \forall g \in G\}$. Orbits do not
intersect and define a partition of $\real^n$. The {\it isotropy
subgroup} (or {\it stabilizer}) $G_x$ of the point $ x\in\real^n$
is the subgroup of $G$ that leaves $x$ unchanged: $G_x = \{g \in G
\mid g \cdot x = x\}$. Isotropy subgroups of points in a same
orbit are conjugated: $G_{g \cdot x} = g \cdot G_x \cdot g^{-1}$.
The {\it orbit type} of the orbit $\Omega(x)$ is the conjugacy
class $[G_x]$ of its isotropy subgroups. The orbit types define a
partition of orbits (and of $\real^n$), each equivalence class
$\Sigma_{[H]}$ is called a {\it stratum} (of $\real^n$):
$\Sigma_{[H]} = \{ x\in\real^n \mid G_x \in [H] \}$. The orbits
and the strata can be partially ordered according to their orbit
types. The orbit type $[H]$ is said to be {\em smaller} than the
orbit type $[K]$: $[H]<[K]$, if $H' \subset K'$ for some $H'\in
[H]$ and $K'\in [K]$. Then $[K]$ is {\em greater} than $[H]$. The
greatest orbit type is $[G]$ and its stratum $\Sigma_{[G]}$
contains all fixed points of the $G$-action. $\Sigma_{[G]}$
coincides with the origin of $\real^n$ if the $G$-action is
effective. Due to the compactness of $G$, the number of different
orbit types is finite and there is a unique smallest orbit type
$[G_p]$, called the {\em principal} orbit type. The stratum
$\Sigma_{[G_p]}$ is called the {\em principal stratum}, all other
strata are called {\em singular}.

The {\em orbit space} is the quotient space $\real^n/G$. The
natural projection $\pi:\real^n \to \real^n/G$ maps orbits of
$\real^n$ into single points of $\real^n/G$. Projections of strata
of $\real^n$ define strata of $\real^n/G$. The principal stratum
of $\real^n/G$ is always open connected and dense in $\real^n/G$.
If $[K]>[H]$, then $\pi(\Sigma_{[K]})$ lies in the boundary of
$\pi(\Sigma_{[H]})$ and the boundary of the principal stratum
contains all singular strata.

A $G$-{\em invariant} function $f:\real^n\to \real$ is such that
$f(g x)=f(x),\ \forall g\in G\,,\ x\in \real^n$. $f$ is then
constant on the orbits and it is in fact a function defined on the
orbit space.

Some symbols. $\real[\real^n]$ and $\real[\real^n]^G$ indicate the
rings of real polynomial functions and of real $G$-invariant
polynomial functions defined in $\real^n$, respectively;
$\real[\real^n]_d$ and $\real[\real^n]^G_d$ indicate the rings of
homogeneous polynomial functions of degree $d$ in $\real[\real^n]$
and in $\real[\real^n]^G$, respectively; $\real[p_1,\ldots,p_q]$
indicates the ring of polynomials with real coefficients in the
$q$ indeterminates $p_1,\ldots,p_q.$

By Hilbert's theorem, $\real[\real^n]^G$ has a finite number $q$
of generators~\cite{springer}: $$\forall\; p\in \real[\real^n]^G,\
\exists\ \widehat p\in \real[p_1,\ldots,p_q]\ \mid \ p(x)=\widehat
p(p_1(x),\ldots,p_q(x)),\ \forall\; x\in \real^n.$$ The $q$
generators for $\real[\real^n]^G$ are called {\em basic
invariants} or {\em basic polynomials} and they form an {\em
integrity basis} for $G$ (and for $\real[\real^n]^G$). The
integrity basis $p_1,\ldots,p_q$ is said to be {\em minimal}
(abbreviated in MIB), if no $p_a,\ a=1,\ldots,q$, is a polynomial
in the other elements of the basis. The basic invariants
$p_1,\ldots,p_q$ may be supposed homogeneous and ordered according
to their degrees $d_a$, for example $d_a\leq d_{a+1}$. One may
choose, in all generality, $p_1(x)=\sum_{i=1}^n x_i^2$. The choice
of a MIB is not unique, but the numbers $q$ and $d_a,
a=1,\ldots,q$ are uniquely determined by $G$.

A MIB can be used to represent the orbits as points of $\real^q$.
In fact, given an orbit $\Omega$, the vector map $p(x)=(p_1(x),
p_2(x),\ldots, p_q(x))\ $is constant on $\Omega$, and, given two
orbits $ \Omega_1 \neq \Omega_2,\ \exists\ p_a\in\mbox{MIB} \mid
p_a(x_1) \neq p_a(x_2),\ \forall x_1\in\Omega_1,\ x_2\in\Omega_2$.
$p(x)$ then defines a point $p=p(x)\in \real^q$ that can be
considered the image in $\real^q$ of $\Omega$ because no other
orbit is represented in $\real^q$ by the same point.\\ The vector
map: $p:\real^n \to \real^q:x \to (p_1(x), p_2(x),\ldots,p_q(x))$,
is called the {\it orbit map} and maps $\real^n$ onto the subset
${\cal S}=p(\real^n)\subset \real^q$. $p$ induces a one to one
correspondence between $\real^n/G$ and ${\cal S}$ so that ${\cal
S}$ can be concretely identified with the orbit space of the
$G$-action. All the strata of ${\cal S}$ are images of the strata
of $\real^n$ through the orbit map and if $[K]>[H]$,
$p(\Sigma_{[K]})$ lies in the boundary of $p(\Sigma_{[H]})$. The
interior of ${\cal S}$ coincides with the image of the principal
stratum and the image of all singular strata lie in the bordering
surface of ${\cal S}$. $p(\Sigma_{[G]})$ always coincides with the
origin of $\real^q$.\\ Generally, the $q$ basic invariants are
algebraic dependent, that is, some polynomial $f\in
\real[p_1,\ldots,p_q]$ exists such that
$f(p_1(x),\ldots,p_q(x))\equiv 0, \forall x\in \real^n$. In this
case the polynomial $f(p_1,\ldots,p_q)$ is called a {\em syzygy
(of the first kind)}. Given a syzygy $f$, the equation
$f(p_1,\ldots,p_q)= 0$ defines a surface in $\real^q$ and ${\cal
S}$ must be contained in that surface. So, ${\cal S}\subset {\cal
Z}$, where ${\cal Z}$ is the intersection of all syzygy
surfaces.\\ All real $G$-invariant $C^\infty$-functions (not only
polynomial functions) can be expressed as real
$C^\infty$-functions of the $q$ basic invariants of a
MIB~\cite{schw-top}, and hence define $C^\infty$ functions on
$\real^q$: $f(x) = \widehat f(p_1(x),\ldots,p_q(x)) \to \widehat
f(p)=\widehat f(p_1,\ldots,p_q)$. The functions $\widehat f(p)$
are defined also in points $p\notin {\cal S}$ but only the
restrictions $\widehat f(p)\mid_{p\in {\cal S}}$ have the same
range as $f(x),\ x \in \real^n$. All $G$-invariant $C^\infty$
functions can then be studied in the orbit space ${\cal S}$ but
one needs to know all equations and inequalities defining ${\cal
S}$ and its strata.

Details and proofs of the statements here recalled may be found
in~\cite{bredon,schw-ihes,mic-zhil,sar-ap} and references therein.

\section{Invariant Theory}\label{Invariant}
Given a MIB $p_1,\ldots,p_q$, there are in general $s\leq q$ basic
polynomials that are algebraic independent:
$p_{h_1},\ldots,p_{h_s}$. If $s<q$ necessarily there are syzygies
(of the first kind). One can determine a finite minimal set of
$r_1$ linearly independent homogeneous polynomials,
$f_1^{(1)},\ldots,f_{r_1}^{(1)}\in\real[p_1,\ldots,p_q]$, of
degrees $m_{11}\leq \ldots \leq m_{1r_1}$ with respect to the
natural grading assigned to the $p_a$ ($\deg(p_a)=\deg(p_a(x))$),
that generate all the syzygies of the first kind as a module over
$\real[p_1,\ldots,p_q]$. One can continue to define syzygies.
Linear combinations of the $r_1$ variables
$f_1^{(1)},\ldots,f_{r_1}^{(1)}$, with coefficients in
$\real[p_{h_1},\ldots,p_{h_s}]$, that vanish identically by
substituting the $f_{i}^{(1)}$ with their expressions in terms of
the $p_a$, are called {\em syzygies of the second kind}. Let
$f_1^{(2)},\ldots,f_{r_2}^{(2)}$ be a linearly independent basis
for the syzygies of second kind, made up by homogeneous
polynomials (with respect to the natural grading) of degrees
$m_{21}\leq \ldots \leq m_{2r_2}$. Linear combinations of the
$r_2$ variables $f_1^{(2)},\ldots,f_{r_2}^{(2)}$, with
coefficients in $\real[p_{h_1},\ldots,p_{h_s}]$, that vanish
identically by substituting the $f_{i}^{(2)}$ with their
expressions in terms of the $f_{j}^{(1)}$, are called {\em
syzygies of the third kind}, and so on. There exist a number $m$
such that one has syzygies of the $m$-th kind
$f_1^{(m)},\ldots,f_{r_m}^{(m)}$, but no syzygies of the
$(m+1)$-th kind. Changing the MIB, one changes the syzygies too
but does not change the numbers $q$, $s$, $m$, $d_a$, $\forall
a=1,\ldots,q$, $r_i$, $m_{ij}$, $\forall\ i=1,\ldots,m$,
$j=1,\ldots,r_i$, that are characteristic of $\real[\real^n]^G$.

By a theorem of Hochster and Roberts, in $\real[\real^n]^G$ one
can choose a maximal subset of $s$ algebraically independent
polynomials: $p_{h_1},\ldots,p_{h_s}$, of degrees
$d_1',\ldots,d_s'$, called {\em primary invariants} (or {\em
homogeneous system of parameters}), and a number $t$ of other
homogeneous polynomials: $f_1,\ldots,f_t$, of degrees
$m_1,\ldots,m_t$, called {\em secondary invariants}, such that all
$p\in \real[\real^n]^G$ can be written in a unique way in
following form:
\begin{equation}p=\sum_{k=0}^{t}q_k\; f_k,
\label{eqn1}\end{equation} where $\ q_k\in
\real[p_{h_1},\ldots,p_{h_s}],$ and, by definition,
$f_0=1$~\cite{stanl,kempf}. In other words, all $p\in
\real[\real^n]^G$ can be written in a unique way in terms of
$p_{h_1}, \ldots, p_{h_s}$, $f_0, \ldots, f_t,$ using the $f_k$ at
most linearly. The most convenient way to choose the $p_{h_l}$ and
the $f_k$ is to take $p_{h_1}, \ldots, p_{h_s}$ and
$f_1,\ldots,f_{q-s},$ equal to the $q$ basic invariants in the
MIB, with $p_{h_1}=p_1$, and $f_{q-s+1},\ldots,f_t,$ equal to some
of the products $f_k f_l$, with $1\leq k\leq l \leq q-s$, in such
a way that all products $f_k f_l$ may be expressed in the form
(\ref{eqn1}) (using in case the syzygies). This choice implies
that the degrees $d_l'$ of the $p_{h_l}$ are equal to $d_a,\  a
\in \{h_1,\ldots,h_s \}$, and the degrees $m_k$ of the $f_k$ are
equal to $d_a, a=1,\ldots,q,\ a \notin \{h_1,\ldots,h_s \}$, or to
their sums. Then, the degrees $m_{1l}$ of the syzygies of the
first kind $f_l^{(1)}$ are $m_{1l}=m_i+m_j$, the degrees $m_{2l}$
of the of the syzygies of the second kind $f_l^{(2)}$ are
$m_{2l}=m_i+m_j+m_k$, and so on.

The main problem of constructive invariant theory is to determine
a MIB for a given group $G$, and after that all the syzygies. It
is then of great interest to find some algorithms that suggest
what the numbers $q$, $s$, $d_a$, $m$, $r_i$, $m_{ij}$, $t$,
$m_k$, are.

A key role is played by the {\em Molien function} $M(\eta)$ that
may be calculated without the knowledge of the MIB in the
following way:\\ $$M(\eta)=\frac{1}{|G|}\;\sum_{g\in
G}\;\frac{1}{\det(e-\eta g)},\quad {\rm or}\quad M(\eta)=\int_{
G}\;\frac{\rmd g}{\det(e-\eta g)},$$ respectively for
finite~\cite{stanl} or compact continuous~\cite{weyl1} groups,
where $\eta<1$ is an abstract variable, $e$ is the unit matrix of
$G$, $|G|$ is the order of $G$, ${\rm d}g$ is the normalized Haar
measure and the integration is over the compact group variety. The
integral may be reduced to a sum of integrals over unit circles in
the complex plane and calculated using the
residues~\cite{stanl1,vp}. An alternative algebraic method to
compute the Molien function for compact connected groups is
presented in~\cite{broer}.

When one calculates the Molien function from the formulas written
above, one ends up with $M(\eta)$ expressed by a rational
function.

The expansion of $M(\eta)$ in power series has the following
interpretation: $$M_1(\eta)=\sum_{d=0}^\infty
\dim(\real[\real^n]^G_d)\eta^d,$$ where
 $\dim(\real[\real^n]^G_d)$ is the dimension
of $\real[\real^n]^G_d$ as a linear space. Let's call $M_1(\eta)$
the {\em first form} of the Molien function. $M_1(\eta)$ gives a
first information about the number and degrees of the basic
invariants of small degrees. One has $\dim(\real[\real^n]^G_d)=
n_d^{(0)}-n_d^{(1)}+n_d^{(2)}-\ldots\;$, where $n_d^{(0)}$ is the
number of linearly independent polynomials of degree $d$,
$n_d^{(1)}$ is the number of linearly independent syzygies of the
first kind of degree $d$, and so on. For sufficiently small $d$,
one may find out the decomposition of $\dim(\real[\real^n]^G_d)$
in terms of $n_d^{(0)}$, $n_d^{(1)}$, \ldots, but in general one
is not able to deduce the numbers $n_d^{(i)}$.

Two rational forms of the Molien function are particularly
interesting.\\ The rational function
$$M_2(\eta)=\frac{\sum_{k=0}^t \eta^{m_k}} {\prod_{l=1}^s
(1-\eta^{d_l'})}\ ,$$ here called the {\em second form} of the
Molien function, has the denominator that is the product of $s$
factors of the kind $(1-\eta^{d_l'})$, one for each primary
invariant $p_{h_l}$, and the numerator that is a reciprocal
polynomial that gives the number $t$ of the secondary invariants
$f_k$, and their degrees $m_k$.\\ The rational function
$$M_3(\eta)=\frac{\sum_{i=0}^m (-1)^i \sum_{j=1}^{r_i}
\eta^{m_{ij}}} {\prod_{a=1}^q (1-\eta^{d_a})}\ ,$$ here called the
{\em third form} of the Molien function, has the denominator that
is the product of $q$ factors of the kind $(1-\eta^{d_a})$, one
for each $p_a$ in a MIB, and the numerator that is a polynomial
that gives a partial information about the number $r_i$ of the
syzygies of $i$-th kind $f_j^{(i)}$, and their degrees $m_{ij}$.

When one calculates $M(\eta)$ one ends up with a rational function
that is generally different from both $M_2(\eta)$ and $M_3(\eta)$.
To get $M_2(\eta)$ and $M_3(\eta)$ from $M(\eta)$ one has to
multiply numerator and denominator of $M(\eta)$ by convenient
factors. Obviously, one has
$M(\eta)=M_1(\eta)=M_2(\eta)=M_3(\eta)$.

If $G$ is finite then $s=n$. If $G$ is a compact Lie group then
$s=n-\dim G+\dim G_p$.

The expressions for $M_2(\eta)$ and $M_3(\eta)$ coincide only if
$q=s$ (and this implies $t=m=0$). Many groups for which $q=s$ are
classified in~\cite{Shep,schw-im,littel,wehlau1}.

 In the general case $s<q$, $M_2(\eta)$ and $M_3(\eta)$ are unknown
as long as one does not know the MIB completely. A possible
expression for $M_2(\eta)$ can easily be found by requiring only
positive coefficients in the numerator. At this point other
possible expressions for $M_2(\eta)$ are obtained by multiplying
the numerator and the denominator of $M_2(\eta)$ by factors like
$(1+\eta^{d'_l})$, where $d'_l$ appears in the exponents of the
denominator, increasing in this way the number of secondary
invariants required by the numerator. Generally, when one starts
to calculate the MIB, one knows the number $s$ of algebraically
independent polynomials, so one is able to write down the correct
form for $M_2(\eta)$. For a given expression of $M_2(\eta)$, one
has to guess the correct expression of $M_3(\eta)$. One notes
that:
\begin{enumerate}
\item
One obtains $M_3(\eta)$ by multiplying numerator and denominator
of $M_2(\eta)$ by a certain (unknown) number $(q-s)$ of factors
like $(1-\eta^{d_a})$, one for each basic non-primary invariant
$f_{k},\ k=1,\ldots,q-s$ (and this implies the equality $m=q-s$).
\item
The coefficients in the numerator of $M_3(\eta)$ corresponding to
the syzygies of the first kind are negative.
\item
An invariant $f_k$, whose degree $m_k$ appears in the numerator of
$M_2(\eta)$, must coincide or with a basic non-primary invariant,
or with a product of them. This implies that: (i) for each $f_k$,
whose degree $m_k$ cannot be written as a sum of degrees of basic
polynomials, there is a factor $(1-\eta^{m_k})$ in the denominator
of $M_3(\eta)$; (ii) the degrees of the syzygies of the first kind
are $m_{1j} \geq 2m_1$.
\end{enumerate}

If one does not know the numbers $q$ and $d_a$, there are many
``possibly right'' expressions for $M_3(\eta)$ that satisfy all
the points written above, but only one is the correct one,
corresponding to the right numbers $q$ and $d_a$. Moreover, given
the right expression for $M_3(\eta)$, the numbers $m$, $r_i$,
$m_{ij}$, are not immediately readable from the numerator because
of simplifications between similar terms. In any case, if there is
only one kind of syzygies of degree $d$ (for example if $d$ is
small), the coefficient of $\eta^d$ in the numerator allows to
know the number of independent syzygies of degree $d$.

Summarizing, the Molien function is a first powerful device to
study $\real[\real^n]^G$ because it gives some information about
the degrees $d_a$ and allows, in any case, to determine correctly
the number and degrees of the basic invariants of small degrees.

When one knows the degrees of some invariants, one may find them
explicitly using known properties of the group action or by means
of the classical procedure of averaging in the group. Given an
arbitrary function $f(x)$, its average in the group is the
invariant function $F(x)$, where: $$F(x)=\frac{1}{|G|}\;\sum_{g\in
G} f(g x),\quad {\rm or}\quad F(x)=\int_{ G}\; f(g x)\;\rmd g, $$
respectively for finite~\cite{burn} or continuous
compact~\cite{weyl} groups. Obviously, for the linearity of $G$,
if $f(x)\in \real[\real^n]_d$, then $F(x)\in \real[\real^n]^G_d$,
so to get an invariant of degree $d$ one may take the average in
the group of a monomial of degree $d$. If $d$ and $n$ are small,
one has only a limited number of possible choices of monomials of
degree $d$, but, if $d$ or $n$ are great, this averaging procedure
becomes difficult to carry out, because of the many possible
monomials of degree $d$ one may choose to average and because most
of the invariants that one finds out in this way are not
independent from those that one already knows.

For a detailed presentation of invariant theory and for the proofs
that are here omitted, the interested reader may read
\cite{springer,stanl,stanl1,vp,sloane,springer1,bens} and the
references therein.\\

\section{$\widehat P$-matrices and Orbit Spaces}\label{PmInOS}

Given a stratum $\Sigma \subset \real^n$, in a point $x\in
\Sigma$, the number of linear independent gradients of the basic
invariants is equal to the dimension of the stratum
$p(\Sigma)\subset {\cal S}$ \cite{sar-jmp}. One may calculate the
$q \times q$ Grammian matrix $P(x)$ with elements $P_{ab}(x)$ that
are scalar products of the gradients of the basic
invariants~\cite{as1}: $$P_{ab}(x) =\nabla p_a(x) \cdot \nabla
p_b(x)=\sum_{i=1}^n \frac{\partial p_a(x)}{\partial
x_i}\frac{\partial p_b(x)}{\partial x_i},$$ so, if $x\in \Sigma$,
$\mbox{rank}(P(x))=\dim (p(\Sigma))$.

From the covariance of the gradients of $G$-invariant functions
($\nabla f(g \cdot x)=g \cdot \nabla f(x)$) and the orthogonality
of $G$, the matrix elements $P_{ab}(x)$ are $G$-invariant
homogeneous polynomial functions of degree $d_a+d_b-2$, so they
may be expressed as polynomials of the basic invariants: $$
P_{ab}(x) =\widehat P_{ab}(p_1(x),\ldots,p_q(x))=\widehat
P_{ab}(p) \quad \forall x \in \real^n\ \mbox{and}\ p=p(x).$$  One
can define then a matrix $\widehat P (p)$ in $\real^q$, called a
{\em $\widehat P$-matrix}, having $\widehat P_{ab} (p)$ for
elements. At the point $p=p(x)\in {\cal S}$, image in $\real^q$ of
the point $x\in\real^n$ through the orbit map, the matrix
$\widehat P(p)$ is the same as the matrix $P(x)$, $\widehat P(p)$
is however defined in all $\real^q$, also outside ${\cal S}$, but
only in ${\cal S}$ it reproduces $P(x),\ \forall x\in\real^n$.\\
$\widehat P(p)$ is  a real, symmetric $q \times q$ matrix and its
matrix elements $\widehat P_{ab}(p)$ are homogeneous polynomial
functions of degree $d_a+d_b-2$ with respect to the natural
grading, $\mbox{rank}(\widehat P(p))$ is equal to the the
dimension of the stratum containing $p$, and ${\cal S}$ is the
{\em only} region of ${\cal Z}$ where $\widehat P(p)$ is positive
semidefinite~\cite{sar-ap,ps}.\\ If $s\leq q$ basic invariants are
algebraic independent, one has $\mbox{rank}(\widehat P(p))=s, \
\forall p$ in the principal stratum, and ${\cal S}$ has dimension
$s$.

The matrix $\widehat P(p)$ completely determines ${\cal S}$ and
its stratification. Defining ${\cal S}_k$ the union of all
$k$-dimensional strata of ${\cal S}$, ($k=1,\ldots,s$), one has:
$${\cal S}=\{p\in {\cal Z} \mid \widehat P(p)\geq 0\},\quad{\cal
S}_k=\{p\in {\cal Z} \mid \widehat P(p)\geq 0,\ \mbox{rank}
(\widehat P(p))=k\}$$ To find out the defining equations and
inequalities of $k$-dimensional strata, one may impose that all
the principal minors of $\widehat P(p)$ of order greater than $k$
are zero and that at least one of those of order $k$, and all
those of order smaller than $k$ are positive.

The \wPm\ contains all information necessary to determine the
geometric structure of the orbit space and its strata, so, if one
wishes to classify geometrically the orbit spaces, it is
sufficient to classify the $\widehat P$-matrices. This, in turn,
gives a non-standard group classification. One discovers so that
different groups, no matter if finite or continuous, share the
same orbit space structure, despite the fact that the strata
correspond to different orbit types (some examples are in Table X
of \cite{st-jmp}). They exist different MIB's (of different linear
groups), with the same number of invariants and the same degrees,
that determine: (i) the same \wPm\ (some examples are in Table X
of \cite{st-jmp}); (ii) different \wPms\ (an example is given by
Entries 6 and 25 of Table V of \cite{st-jmp}).

Some examples using the \wPms\ to study orbit space
stratifications, minima of invariant polynomials and phase
transitions may be found in \cite{sv,st-jmp,gpstvv-jmp}.

\section{$\widehat P$-matrices and Invariant Theory}\label{PmInIT}

The {\em \wPm\ method} to determine a MIB is summarized in the
following items (i)-(iv) and applied in three examples.

Let $p_1,\ldots,p_q$ be a set of $q$ independent $G$-invariant
homogeneous polynomials.

\begin{enumerate}
\item \label{uno} With $p_1,\ldots,p_q$ one tries to construct the \wPm.
By Euler's theorem on homogeneous functions and the standard
choice for $p_1$, one has $\widehat P_{1a}(p)=\widehat P_{a1}(p)=2
d_a p_a,\ \forall a=1,\ldots,q$. Taking into account the symmetry
of $\widehat P(p)$, one has then to calculate only the matrix
elements $\widehat P_{ab}(p)$, with $2\leq a \leq b \leq q$.
\item \label{due} If it is not possible to express $P_{ab}(x)$ as a polynomial
in $p_1(x),\ldots,p_q(x)$, then $P_{ab}(x)$ is an invariant of
degree $d_a+d_b-2$ that is independent from the known ones. One
has then to add a new basic invariant to the set $p_1,\ldots,p_q$.
One may choose, for example, an irreducible invariant factor of
$P_{ab}(x)$ independent from $p_1,\ldots,p_q$.

\item \label{tre} When one succeeds to express all the matrix
elements $P_{ab}(x)$ in term of the basic invariants, one has
found a ``closed'' \wPm\ $\widehat P(p)$ and a set
$p_1,\ldots,p_q$ that is a MIB (or part of it) for a group
$\widehat G$ that either coincides with $G$ or contains $G$ as a
subgroup (see Example 3).

\item \label{quattro} To check if $p_1,\ldots,p_q$ do form a complete MIB for
$G$ one has to look to the Molien function of $G$. If $M_2(\eta)$
and  $M_3(\eta)$ agree with the MIB $p_1,\ldots,p_q$, then this is
the MIB for $G$ one was searching in. If instead $M_3(\eta)$
requires a different set of degrees for the basic polynomials,
then one has to find out (by averaging in the group or in some
other way) at least one of the missing basic polynomials of the
lowest degree. If $p_0$ is this new invariant, first of all one
has to check if the higher degree invariants in the set
$p_1,\ldots,p_q$ are all independent from $p_0$, otherwise one has
to discard the non independent ones. Then, return to item
(\ref{uno}) above with $p_0$ in the set of the $q$ invariant
polynomials.
\end{enumerate}
Some technical suggestions. With the \wPm\ method, with only
invariants of degree 2 one is not able to determine higher degrees
invariants and with only even degree invariants one is not able to
determine odd degree invariants. So it is convenient to start to
build the \wPm\ having determined in some way at least all
invariants of degree 2, all invariants of the smallest odd degree
and all the invariants of the smallest even degree greater than 2.
One then starts to determine the \wPm\ as indicated in items
(\ref{uno})-(\ref{quattro}) above, proceeding step after step to
calculate the matrix elements of increasing degrees. During this
calculation, every time one adds a basic invariant to the set of
basic invariants $p_1,\ldots,p_q$, according to item (\ref{due})
above, it is convenient to look if this MIB is consistent with the
guessed expression of $M_2(\eta)$ and $M_3(\eta)$. In case, one
has to modify the expressions of $M_2(\eta)$ and/or $M_3(\eta)$ to
include the new basic invariant. One has so a partial check of the
correctness either of $M_2(\eta)$ and $M_3(\eta)$, either of the
MIB. Every time one starts to determine the matrix elements
$\widehat P_{ab}(p)$ of a given degree $d$, it is convenient to
control if one knows explicitly all the basic invariants of degree
lower than $d$ that are predicted by $M_3(\eta)$. If one misses
one of these invariants it is strongly recommended to find out
this basic invariant first, otherwise, one does longer and harder
calculations.

If a subset $p_{l_1},\ldots,p_{l_k}$ of the MIB for $G$ is a
complete MIB for a group $\widehat G \subseteq O(n,\real)$,
containing $G$ as a subgroup, then the principal minor of the
\wPm\ of $G$ formed by the rows and columns $l_1,\ldots,l_k$ is
expressed only in terms of
$p_{l_1},\ldots,p_{l_k}$.\\

\noindent Example 1. Let $G=O_h$, the symmetry group of the cube.
The Molien function of $G$ has the following expression: $$
M_2(\eta)=M_3(\eta)=\frac{1}{(1-\eta^2) (1-\eta^4)(1-\eta^6)} $$
and predicts a MIB with 3 algebraic independent polynomials of
degree 2, 4 and 6 that have been calculated by many authors (see
for ex. Table 4 of \cite{mic-zhil}). As an elementary example of
the \wPm\ method to determine a MIB, let us start with the basic
invariants $p_1=x^2+y^2+z^2$ and $p_2=x^4+y^4+z^4$. The only non
trivial element of the \wPm\ formed with $p_1$ and $p_2$ is
$\widehat P_{22}$. One finds easily $\nabla p_2=(4x^3,4y^3,4z^3)$
and $ P_{22}=\nabla p_2\cdot \nabla p_2=16(x^6+y^6+z^6)$. As
$\widehat P_{22}$ cannot be written in the form $c_1 p_1^3+c_2 p_1
p_2$, the basis $p_1$ and $p_2$ is not complete and one may take
$p_3=x^6+y^6+z^6$ for a third element of the MIB. All elements of
the \wPm\ constructed with $p_1$, $p_2$ and $p_3$ can be written
only in terms of $p_1$, $p_2$ and $p_3$, so the MIB is complete,
it agrees with the Molien function and in fact is a possible MIB
for $G$. This trivial Example shows that by constructing the \wPm\
with only part of the MIB known, one succeeds to find out all the
missing basic invariants.\\

\noindent Example 2. Let $G$ be the group of order 192 generated
by the matrices $$g_1=\frac{\sqrt{2}}{2}\left(\begin{array}{cccc}
1&1&0&0\\ 1&-1&0&0\\ 0&0&1&1\\0&0&1&-1\\
\end{array}\right),\qquad
g_2=\left(\begin{array}{cccc} 1&0&0&0\\ 0&0&0&-1\\
0&0&1&0\\0&1&0&0\\
\end{array}\right).$$
This group is the realification of the complex unitary group
considered in \cite{sloane}, and one is referred to \cite{sloane}
for details about its structure.

The Molien function of $G$ has the following first and second
form:
 $$  M_1(\eta)=
1+\eta^2+\eta^4+\eta^6+4\eta^8+4\eta^{10}+7\eta^{12}+
7\eta^{14}+12\eta^{16}+$$ $$ +
17\eta^{18}+22\eta^{20}+22\eta^{22}+ 36\eta^{24}+ 43\eta^{26}+
\ldots
 , $$$$  M_2(\eta)=\frac{1+2\eta^{8}+
2\eta^{12}+2\eta^{16}+5\eta^{18}+5\eta^{24}+2
\eta^{26}+2\eta^{30}+2\eta^{34}+\eta^{42}}{(1-\eta^2)
(1-\eta^8)(1-\eta^{12})(1-\eta^{24})}\;.$$ From $M_1(\eta)$ one
gets the following kind of information. The term $4\eta^8$
suggests that there are 4 lin. indep. invariants of degree 8:
$p_1^4$, $p_2$, $p_3$, $p_4$. The term $7\eta^{12}$ suggests that
there are 7 lin. indep. invariants of degree 12: $p_1^6$, $p_1^2
p_2$, $p_1^2 p_3$, $p_1^2 p_4$, $p_5$, $p_6$, $p_7$. The term
$12\eta^{16}$ suggests that there are 12 lin. indep. invariants of
degree 16: 13 lin. indep. invariants are the terms in the
expansion of the polynomial $(p_2+p_3+p_4+p_1^4)^2+(p_5+p_6+
p_7)p_1^2$ and it must exist one syzygy of the first kind of
degree 16, $f_1^{(1)}$, so that $12=13-1$. The term $17\eta^{18}$
suggests that there are 17 lin. indep. invariants of degree 18: 12
independent invariants are found by multiplying by $p_1$ the 12
independent invariants of degree 16, and there must exist 5 new
basic invariants of degree 18: $p_8$, $p_9$, $p_{10}$, $p_{11}$,
$p_{12}$, so that $17=12+5$. One may continue this analysis and
find out that there are 4 syzygies of the first kind of degree 20:
$f_2^{(1)},\ldots,f_5^{(1)}$ and no basic invariants and no
syzygies of degree 22. The analysis becomes very uncertain for
degree 24 because one might have either basic invariants, either
syzygies of the first and of the second kind, and one only knows
that their numbers $n_{24}^{(0)}$, $n_{24}^{(1)}$ and
$n_{24}^{(2)}$ must satisfy the condition
$n_{24}^{(0)}-n_{24}^{(1)}+n_{24}^{(2)}=36$.

The information obtained so far from $M_1(\eta)$ and $M_2(\eta)$
suggest to obtain the third form of the Molien function by
multiplying numerator and denominator of $M_2(\eta)$ by
$(1-\eta^8)^2(1-\eta^{12})^2(1-\eta^{18})^5$. Note that the
exponents 2, 2, 5 are the coefficients of $\eta^8$, $\eta^{12}$,
$\eta^{18}$ in the numerator of the second form of the Molien
function. Doing this, one finds a third form of the Molien
function that shows at the numerator the existence of the syzygies
$f_1^{(1)},\ldots,f_5^{(1)}$ but suggests, wrongly, that there is
only one basic invariant of degree 24. At this point it is
convenient to stop the (uncertain) analysis of the Molien function
and start to build the \wPm.

The Molien function suggests so far a MIB with 1 invariant of
degree 2: $p_1$, 3 invariants of degree 8: $p_2, p_3, p_4$, 3
invariants of degree 12: $p_5, p_6, p_7$, 5 invariants of degree
18: $p_8, p_9, p_{10}, p_{11}, p_{12}$, and one invariant of
degree 24: $p_{13}$. In this example the explicit calculation of
all the invariants by the method of averaging in the group is not
trivial, because of the high degree of the invariants, so, it is
convenient to use the \wPm\ method.

First of all, one defines: $p_1(x)=x_1^2+x_2^2+x_3^2+x_4^2$ and
finds out the three invariants of degree 8, for example by
averaging in the group the monomials $x_1^2 x_2^2 x_3^2 x_4^2$,
$x_1^4 x_3^4$ and $x_1^5 x_3^3$. One then starts to build up the
\wPm. One calculates first the matrix elements of degree 14: $
P_{2,2}(x),\ldots,P_{4,4}(x)$ and one finds out the 3 basic
invariants of degree 12. One then calculates the matrix elements
of degree 18: $P_{2,5}(x),\ldots, P_{4,7}(x)$, and one finds out
the 5 basic invariants of degree 18. One then calculates the
matrix elements of degree 22: $P_{5,5}(x),\ldots, P_{7,7}(x)$,
with no need of new basic invariants. One then calculates the
matrix elements of degree 24: $P_{2,8}(x),\ldots, P_{4,12}(x)$,
and one finds out two basic invariants of degree 24: $p_{13}(x)$
and $p_{14}(x)$, not only one, as expected. Then the third form of
the Molien function has to be corrected by multiplying its
numerator and denominator by $(1-\eta^{24})$ and obtaining:
$$M_3(\eta)= \frac{1-\eta^{16}-4 \eta^{20}- \eta^{24} -8
\eta^{26}+4 \eta^{28}-8 \eta^{30}+8 \eta^{34}-19 \eta^{36}+\ldots}
{(1-\eta^2) (1-\eta^8)^3 (1-\eta^{12})^3 (1-\eta^{18})^5
(1-\eta^{24})^2}\;.$$\\

Calculating all the remaining elements of the \wPm\ one does not
need any new basic invariant. As $p_1,\ldots,p_{14}$ agrees with
the third form of the Molien function, they form a complete MIB
for $G$. It is important to note that the invariant $p_{14}$ was
not predicted from the Molien function, but it is required to
express some of the elements of the matrix $P(x)$ in terms of the
MIB.

This example shows that the Molien function is not sufficient to
specify the number and degrees of all the basic invariants in a
MIB, but its combined use with the construction procedure of the
\wPm\ allows one to do that. Moreover, the \wPm\ method is
constructive, in the sense that it allows to determine concretely
the MIB.

In this case no proper subset of the MIB $p_1,\ldots,p_{14}$, is a
complete MIB for a group $\widehat G$ containing $G$ as a
subgroup, because no principal minor of the \wPm\ can be written
only in terms of the corresponding subset of basic invariants.

Having found the MIB, all the syzygies of any kind can be
determined.\\

\noindent Example 3. Let $G$ be the cyclic group of order 3,
generated by the matrix
$$g=\left(\begin{array}{cc} -\frac12 & -\frac{\sqrt{3}}{2} \\
\frac{\sqrt{3}}{2}  & -\frac12 \\
\end{array}\right).$$
The three forms of the Molien function are easily found. The
second form is
$$M_2(\eta)=\frac{1+\eta^3}{(1-\eta^2)(1-\eta^3)}$$
and suggests a MIB of 3 elements of degree 2, 3, 3. Using the
standard form of $p_1$ and averaging the monomials $x_1^3$ and
$x_2^3$, one finds out the following MIB:
$$ p_1=x_1^2+x_2^2,\quad p_2=x_1^3-3 x_1 x_2^2, \quad
p_3=x_2^3-3 x_2 x_1^2\; . $$ If one uses the \wPm\ method in this
trivial example starting with the basic invariants $p_1$ and
$p_2$, one does not need any other invariant to write down the
\wPm\ and would not find $p_3$. The group that leaves $p_1$ and
$p_2$ invariant however is $G$, that has 3 invariants, according
to its Molien function. To use the \wPm\ method it is then
necessary to start with all invariants of the smaller degree
greater than 2.

Let now $H$ be the cyclic group of order 3, generated by the
$4\times 4$ matrix
$$h=\left(\begin{array}{cc} g & e\\
e & g\\
\end{array}\right),$$
where  $e$ is the unit matrix of order 2 and $g$ is the matrix
written above. The three forms of the Molien function are easily
found. The second form is
 $$ M_2(\eta)=\frac{1+2 \eta^2+6
\eta^3+2\eta^4+ \eta^6 }{(1-\eta^2)^2 (1-\eta^3)^2}\; ,$$ and
suggests a MIB of 12 elements: $p_1,\ldots,p_{4}$ of degree 2, and
$p_5,\ldots,p_{12}$ of degree 3. All the basic invariants can be
easily found by averaging in the group. For example, one may use
the standard form of $p_1$ and the averages of the monomials
$x_1^2$, $x_1 x_3$, $x_1 x_4$, $x_1^3$, $x_2^3$, $x_3^3$, $x_4^3$,
$x_1 x_2 x_3$, $x_1 x_2 x_4 $, $x_1 x_3 x_4 $, $x_2 x_3 x_4 $ for
the invariants $p_2,\ldots,p_{12}$. One verifies easily that none
of them can be expressed as a polynomial in the other ones, so
they form a MIB for $H$.

In this case if one uses the \wPm\ method starting with
$p_1,\ldots,p_4$ and one arbitrary invariant of degree 3 one would
also find out all the basic polynomials of a MIB (generally with
more complicated expressions than those obtained from the
averages).

If one uses the \wPm\ method starting for example with the basic
invariants $p_1=x_1^2+x_2^2+x_3^2+x_4^2$, $p_2=x_1^2+x_2^2$ and
$p_5=x_1^3-3x_1 x_2^2$, one does not need any other invariant to
write down the \wPm. One knows from the Molien function that the
polynomials $p_1,p_2,p_5$ do not form a MIB for $H$ and one
understands that one has to use more invariants at the beginning
to find out the MIB with the \wPm\ method. \\The invariance group
of $p_1,p_2,p_5$ is the group $\widehat H$, containing $H$ as a
subgroup, generated by an arbitrary rotation in the plane
$x_3,x_4$ and by the matrix $g$ acting in the plane $x_1,x_2$. We
know from above, or from the Molien function of $\widehat H$:
$$M_2(\eta)=\frac{1+\eta^3}{(1-\eta^2)^2(1-\eta^3)}\; ,$$
that a MIB for $\widehat H $ has 2 invariants of degree 2 and 2 of
degree 3, so $p_1,p_2,p_5$ is not a complete MIB even for
$\widehat H$.

This example suggests that it is convenient to start using the
\wPm\ method after having determined all invariants of degree 2
and all invariants of the smaller degree greater than 2 predicted
by the Molien function. Moreover, it shows that the Molien
function plays a fundamental role to decide if a given set of
invariants do form a complete MIB for some group or not.


\section*{References}
\begin{thebibliography}{10}
\bibitem{springer} Springer T A 1977 {\it Invariant Theory}
 ({\it Lecture Notes in Mathematics} vol~585) (Berlin: Springer-Verlag)

\bibitem{schw-top} Schwarz G W 1975 Smooth functions invariant under the
action of a compact Lie group {\it Topology} {\bf 14} 63--8

\bibitem{bredon} Bredon G 1972 {\it Introduction to Compact Transformation Groups} (New York: Academic Press)

\bibitem{schw-ihes} Schwarz G W 1980 Lifting smooth homotopies of orbit spaces {\it I.H.E.S. Publ. Math.} {\bf 51} 37--135

\bibitem{mic-zhil} Michel L and Zhilinskii B I 2001 Symmetry, invariants, topology. Basic tools {\it Phys.
Rep.} {\bf 341} 11--84

\bibitem{sar-ap} Abud M and Sartori G 1983 The geometry of spontaneous symmetry breaking
 \APNY {\bf 150} 307--72

\bibitem{stanl} Stanley R P 1979 Invariants of finite groups and their
applications to combinatorics {\it Bull. Amer. Math. Soc.} {\bf 1}
475--511

\bibitem{kempf} Kempf G 1979 The Hochster-Roberts theorem of invariant
theory {\it Michigan Math. J.} {\bf 26} 19--32

\bibitem{weyl1} Weyl H 1926 Zur Darstellungstheorie und Invariantenabz\"ahlung
der projektiven, der Komplex und der Drehungsgruppe {\it Acta
Mathematica} {\bf 48} 255--78, reprinted in 1968 {\it Ges. Abh.
Bd. III} (Berlin: Springer-Verlag) pp~1--25

\bibitem{stanl1} Stanley R P 1979 Combinatorics and invariant theory
{\it Proc. of Symp. in Pure Mathematics} vol~34 ed D K
Ray-Chaudhuri (Providence RI: American Mathematical Society)
pp~345--55

\bibitem{vp} Popov V L and  Vinberg E B 1994 Invariant theory  {\it
Algebraic Geometry IV} ({\it Encyclopaedia of Mathematical
Sciences} vol~55)  ed A N Parshin and I R Shafarevich (Berlin:
Springer-Verlag) pp~127--278

\bibitem{broer} Broer B 1994 A new method for calculating Hilbert
series {\it J. Algebra} {\bf 168} 43--70


\bibitem{Shep} Shephard G C Todd J A 1954 Finite unitary reflection
groups {\it Canad. J. Math.} {\bf 6}  274--304

\bibitem{schw-im} Schwarz G W 1978 Representations of simple Lie groups with
 regular rings of invariants {\it Invent. Math.} {\bf 49}  167--91

\bibitem{littel} Littelmann P 1989 Koregul\"are und \"aquidimensionale
Darstellungen {\it J. Algebra} {\bf 123}  193--222

\bibitem{wehlau1} Wehlau D L 1993 Equidimensional representations of 2-simple groups
{\it J. Algebra} {\bf 154}  437--89

\bibitem{burn} Burnside W 1911 {\it Theory of Groups of Finite Order} second
ed. (Cambridge: Cambridge University Press)

\bibitem{weyl} Weyl H 1939 {\it The Classical Groups} (Princeton: Princeton University Press)

\bibitem{sloane} Sloane N J A 1977 Error-correcting codes and invariant theory:
new applications of a nineteenth-century technique {\it Amer.
Math. Monthly} {\bf 84} 82--107

\bibitem{springer1} Springer T A 1982 S\'eries de Poincar\'e
 dans la theorie des invariants {\it S\'eminaire
 d'Alg\`ebre}
({\it Lecture Notes in Mathematics} vol~1029) (Berlin:
Springer-Verlag) pp~37--54

\bibitem{bens} Benson D J 1993 {\it Polynomial Invariants of Finite Groups} ({\it London
Mathematical Society Lecture Notes Series} vol~190) (Cambridge:
Cambridge University Press)

\bibitem{sar-jmp} Sartori G 1983 A theorem on orbit structure (strata) of compact
 linear Lie groups \JMP {\bf 24}  765--8

\bibitem{as1} Abud M and Sartori G 1981 The geometry of orbit space and natural minima of Higgs potentials \PL {\bf 104 B} 147--52

\bibitem{ps} Procesi C and Schwarz G W 1985 Inequalities defining orbit
spaces {\it Invent. Math.} {\bf 81}  539--54

\bibitem{st-jmp} Sartori G and Talamini V 1998 Orbit spaces of compact coregular
simple Lie groups with 2, 3 and 4 basic polynomial invariants:
effective tools in the analysis of invariant potentials \JMP {\bf
39} 2367--401

\bibitem{sv} Sartori G and Valente G 1996 Orbit spaces of reflection groups with 2, 3 and 4 basic polynomial invariants, \JPA {\bf 29}
193--223

\bibitem{gpstvv-jmp} Gufan Yu M, Popov Al V, Sartori G, Talamini V, Valente G and Vinberg E B 2001
Geometric invariant theory approach to the determination of ground
states of D-wave condensates in isotropic space \JMP {\bf 42}
1533--62

\end{thebibliography}
\end{document}